# Molecular motors robustly drive active gels to a critically connected state

José Alvarado[1], Michael Sheinman[2], Abhinav Sharma[2], Fred C. MacKintosh[2*], Gijsje H. Koenderink[1*]

[1] FOM Institute AMOLF, Science Park 104, 1098 XG Amsterdam, The Netherlands
[2] Department of Physics and Astronomy, Vrije Universiteit, 1081 HV Amsterdam, The Netherlands
* e-mail:fcm@nat.vu.nl, gkoenderink@amolf.nl

## **Abstract**

Living systems often exhibit *internal driving*: active, molecular processes drive nonequilibrium phenomena such as metabolism or migration. Active gels constitute a fascinating class of internally driven matter, where molecular motors exert localized stresses inside polymer networks. There is evidence that network crosslinking is required to allow motors to induce macroscopic contraction. Yet a quantitative understanding of how network connectivity enables contraction is lacking. Here we show experimentally that myosin motors contract crosslinked actin polymer networks to clusters with a scale-free size distribution. This critical behavior occurs over an unexpectedly broad range of crosslink concentrations. To understand this robustness, we develop a quantitative model of contractile networks that takes into account network restructuring: motors reduce connectivity by forcing crosslinks to unbind. Paradoxically, to coordinate global contractions, motor activity should be low. Otherwise, motors drive initially well-connected networks to a critical state where ruptures form across the entire network.

**Introduction**

One of the defining qualities of soft matter is that it is readily driven far from thermodynamic equilibrium by external stress. Driving forces such as those due to an electric field or shear can drive colloidal suspensions and polymer networks into fascinating non-equilibrium patterns, including banded[1,2], jammed[3], and randomized steady states[4]. Much progress has been made in understanding such *externally* driven systems[5]. By contrast, living soft matter systems such as cells and tissues naturally exhibit a unique form of *internal* driving in the form of mechanochemical activity[6,7]. A prominent example is the *cytoskeleton*, a meshwork of protein polymers and force-generating motor proteins that constitutes the scaffold of cells. In solutions of purified cytoskeletal filaments and motors, remarkable self-organized patterns have been observed[8,9], inspiring theoretical work of these so-called active gels[10].

More recently, attention has shifted to the important role of network connectivity in active gels, which can be controlled by the number of crosslinks between filaments. In weakly connected systems, motors slide filaments to form static or dynamic clusters[11-14]. In the opposite limit of a well-connected, elastic network, motors generate contractile stresses as they pull against crosslinks, which can dramatically change the elastic properties of the network[15,16] or lead to contraction[17,18]. The existence of a threshold connectivity that separates these two behaviors has been proposed, since macroscopic contractions are known to occur above certain minimum values of crosslink or actin concentration[14,17,19,20]. We should expect remarkable critical behavior at the threshold of contraction. Recent theoretical models predict diverging correlation length-scales and a strong response to external fields[21-24] at the threshold of rigidity. In suspensions of self-propelled patches, critical slowing was predicted at the threshold of alignment[25]. Yet the threshold of contraction still remains poorly understood, and experimental evidence of criticality in active gels remains lacking.

Here, we experimentally study model cytoskeletal systems composed of actin filaments and myosin motors. We vary network connectivity over a broad range by adding controlled amounts of crosslink protein. We show that the motors can actively contract the networks into disjoint clusters that exhibit a power-law size distribution. This behavior is reminiscent of classical conductivity percolation[26], for which a power-law size distribution of clusters occurs close to a critical point. However, in sharp contrast to this equilibrium phenomenon, we observe critical behavior over a wide range of initial network connectivities. To understand this robustness, we develop a general theoretical model of contractile gels that can quantitatively account for our observations. In this model, motors not only contract the network, but also reduce the connectivity of initially stable networks down to a marginal structure by promoting crosslink unbinding. Below this marginal connectivity, the network no longer supports stress and the system rapidly devolves to disjoint clusters which reflect the critical behavior of the marginal structure. Our model predicts cluster size distributions that agree well with experiment. Moreover, it predicts an inverse relationship between cluster size and motor activity, which we also confirm experimentally.

**Experiment: motors rupture networks into clusters**

In order to resolve the interplay between motor activity and network connectivity in active cytoskeletal networks, we develop a biomimetic model system with a well-controlled composition (Fig 1a). Networks are formed by initiating actin filament polymerization, which results in a semiflexible polymer meshwork with a pore size of ~0.3 μm. We control the motor activity by adding different amounts of myosin motors, expressed in terms of the myosin-to-actin molar ratio, $R_M$ = [myosin] / [actin]. We control the network connectivity by adding different amounts of the crosslink fascin, which can simultaneously bind to two neighboring actin filaments (see Methods). We express the crosslink density in terms of the fascin-to-actin molar ratio, $R_C$ = [fascin] / [actin]. To ensure that we can observe motor-driven contraction on all scales, from microscopic to macroscopic, we prepare networks in customized flow-cells, which fit entirely in the field-of-view of the 4× objective of a confocal microscope (see Methods). To track the temporal evolution of the networks, we acquire time-lapse movies starting from 1 minute after the initiation of actin polymerization, where the solution is still homogeneous, until 2 hours afterwards.

To resolve the influence of network connectivity, we first prepare a series of networks with constant myosin activity ($R_M$ = 0.01) and gradually increasing crosslink density ($R_C$). Even at low $R_C$, the motors can contract actin networks (Supplementary Movie 1). However, contraction occurs only on a small length scale, as seen in the time projection image in Fig. 1b. However, when we increase $R_C$, contraction occurs on a larger length scale (Fig 1c, Supplementary Movie 2). The motors break the network up into multiple disjoint clusters. At still higher $R_C$, motor activity contracts the entire network into a single dense cluster which often retains the square shape of the assay chamber (Fig 1d, Supplementary Movie 3).

To quantify the effect of connectivity on the length scale of network contraction, we developed an image processing algorithm (Supplementary Movie 4) which identifies the clusters in the final image and traces their origin back in time. As shown in Fig. 1, the initial areas of each cluster are small in weakly crosslinked networks (panel d). The smallest clusters are ~30 μm in size, which corresponds to the typical distance between myosin motor clusters in the absence of cross-links (Supplementary Figure 1). However, the clusters increase in size when the crosslink density is increased (panel f). In strongly crosslinked networks, the entire network forms one cluster (panel g).

Qualitatively, the transition from local to macroscopic contraction is reminiscent of a classical conductivity percolation transition. Below this transition, a system is only locally correlated and cannot establish connections over long distances. Only above a certain critical connectivity can the system establish global correlations. In order to determine the extent of agreement between our experimental results and percolation theory, we investigate three key predictions[26].

First, conductivity percolation theory predicts how connectivity determines the size of the largest and second-largest connected clusters. Connectivity is quantified by the probability $p$ of creating a connection. The largest cluster (of size $\xi_1$) is predicted to increase monotonically with $p$, while the second-largest cluster (of size $\xi_2$) should exhibit a peak right at the conductivity percolation threshold, where $\xi_1$ and $\xi_2$ both approach the system size, $L$ (Fig 1h, inset). Our experiments agree with this prediction: the measured cluster sizes, $\xi_1$ and $\xi_2$, are both small at low crosslink density and increase monotonically with increasing crosslink concentration until they approach the system size, $L \approx 2.5$ mm, around $R_C \sim 0.01$ (Fig 1h). Above this threshold connectivity, $\xi_1$ remains close to $L$ whereas $\xi_2$ decreases towards zero as the entire network contracts to one large cluster.

Second, percolation theory predicts how cluster sizes are distributed: around the critical point, we should find a power law with an exponent of −2. To test this prediction, we begin by looking for networks which satisfy $\xi_1 \sim \xi_2 \sim L$. We replot all measurements separately in $\xi_1$-$\xi_2$-space (Fig. 2a). Because $\xi_2 < \xi_1$ by definition, all samples are located within a triangle in $\xi_1$-$\xi_2$-space. We can clearly identify the samples at the triangle's peak, where $\xi_1 \sim \xi_2 \sim L$. We denote this peak as the *critically connected regime*. To the left of the peak are samples with low $R_C$, which we denote the *local contraction regime*. To the right of the peak are samples with high $R_C$, the *global contraction regime*.

Do the samples in the critically connected regime really exhibit critical behavior? To test this more rigorously, we plot the entire distribution of cluster sizes (Fig 2b). We represent the observed distribution as a histogram (open circles), where power-law distributions appear as straight lines on a log-log plot. We additionally plot complementary cumulative probability distributions (solid lines), whose visual form does not depend on bin size. We find that our experiments are again consistent with percolation theory: the critically-connected regime indeed exhibits a cluster-size distribution that is statistically consistent with a power-law across more than two orders of magnitude in measured area[27]. The power-law exponent is −1.9, close to the exponent of −2 predicted by percolation theory. The distributions of the other two regimes furthermore agree with percolation theory. The local contraction regime exhibits a short-tail distribution with a sharp cut-off. The global contraction regime exhibits a bimodal distribution with two well-separated length scales: the percolating cluster with size $\xi_1 \sim L$ and other small disjointed clusters with a typical size of $\xi_2 \ll L$.

Third, percolation theory predicts that only systems that are close to the critical point should exhibit a power law. But this prediction is difficult to reconcile with our data: the critically connected regime in $\xi_1$-$\xi_2$-space (Fig 2a) is populated by samples which span a wide range of cross-link densities (from $R_C = 0.01$ to $R_C = 0.1$). This is also reflected in Fig. 1h, which shows a broad $\xi_2$-peak that is over half an order of magnitude wide in $R_C$, in sharp contrast with the narrow $\xi_2$-peak expected from percolation theory (inset of Fig 1h). We can therefore conclude that classical conductivity percolation theory cannot provide a complete description of the physics of active, contractile networks.

## Simulation: network restructuring

Percolation theory describes a network with a fixed connectivity. This can be appropriate for equilibrium fiber networks without internal driving. However, in motor-driven networks, the total connectivity can change significantly[28-30]. When we image our networks at high resolution, we see that motors actively pull on network strands and disconnect them, thereby reducing connectivity (Supplementary Movie 5). Crosslinks bind only transiently (~10 s in case of fascin[31]), and their binding kinetics are typically stress-dependent[32]. There is strong evidence that unbinding of fascin crosslinks is promoted under stress. For instance, in gliding assays where actin-fascin bundles move over immobilized myosin motors, the motors actively zipped open the bundles[33]. We hypothesize that such stress-dependent binding kinetics allow motor activity to drive initially well-connected networks down towards a critically connected state.

To test this hypothesis, we develop a computational model of contractile actin-myosin networks using molecular dynamics. We model actin filaments with a planar triangular lattice of nodes connected by line segments of length $l_0$ (Fig 3a). Filaments possess stretching modulus $k$ and can strain-stiffen[34] and buckle[35]. We set the average number $z$ of line segments connected to a node (i.e. coordination number) to 4.0. Point-like crosslinks are randomly placed on nodes with probability $p$, which depends on crosslink concentration $c$. We assume first-order kinetics of crosslink (un)binding, which yields $p = c/(1+c)$. We model the crosslinks by freely-hinged constraints, which prevent relative sliding of connected filaments. Motor activity results in contractile stresses[13,36,37], which we model by pairs of forces $f$ between nodes. Every node has mobility $\mu$ and experiences an effective, free-draining viscosity, $\eta$. The network evolves over time to achieve force balance at the nodes (Fig 3b). For fixed crosslinks, network connectivity remains unchanged and $\xi_1$ and $\xi_2$ remain constant. We now introduce into the model the important ingredient of *network restructuring*: connectivity can change via crosslink unbinding and rebinding. The unbinding rate of a crosslink $k_{off}$ increases exponentially with the tension $T$ according to Bell's law[32]: $k_{off} = k_{off,0} \exp(T / T_0)$, where $k_{off,0}$ denotes the off-rate in the absence of tension, and $T_0$ a characteristic tension (Fig 3c). To account for rebinding, we consider the probability that an unbinding event is followed by a rebinding event at the same location before filaments are separated, which is given by $\exp(-c\, k_{on}\, d / T\, \mu)$, where $d$ is an effective distance on the order of the mesh size over which filaments can move with velocity equal to $T\, \mu$ and $k_{on}$ is the binding rate of a crosslink. The effective unbinding rate is thus given by

$$k_{off} = k_{off,0} \exp(T / T_0) \exp(-c\, k_{on}\, d / T\, \mu).$$

By varying $c$ across many simulations (keeping $f$ constant), we recover the three regimes found in experiment: the local contraction (Fig 3d,e; Supplementary Movie 6), critically connected (Fig 3f,g; Supplementary Movie 7), and global contraction regimes (Fig h,i; Supplementary Movie 8). The crosslink-dependence of $\xi_1$ and $\xi_2$ versus $c$ (Fig 3j) as well as the cluster size distributions (Supplementary Figure 2) are fully consistent with experiment. The model clearly reveals that motor activity broadens the $\xi_2$-peak: in the absence of active network restructuring (panel j, open symbols), only a narrow region (yellow stripes) around the critical point exhibits critical behavior. In the presence of network restructuring (panel j, closed symbols), this region broadens (solid yellow box). Motor-driven network restructuring can therefore account for the surprising robustness of critical behavior we found in experiment.

**Motors promote network restructuring**

So far we have investigated the effect of connectivity in experiment and simulation ($R_C$ and $N$), but kept motor activity constant ($R_M$ and $f$). Network restructuring breaks networks into clusters because motor stresses unbind crosslinks. Increased motor activity should therefore increase tension on crosslinks, enhance their unbinding, and lead to smaller clusters. To test this hypothesis, we simulate well-connected networks with constant $c$ but varying motor activity (modeled through changes in the force $f$). Increased force indeed leads to smaller clusters (Fig 4a-d). At low force ($f/k < 0.1$), the networks contract macroscopically ($\xi_1 \gg \xi_2$), while at higher force levels $\xi_1$ sharply decreases as f increases (Fig 4e). Increasing $f$ allows networks that would otherwise contract globally to exhibit clusters with a power-law size distribution.

In order to validate these predictions, we perform experiments in well-connected networks ($R_C = 0.02$) where we change motor activity by varying the myosin-to-actin molar ratio $R_M$. In agreement with the model's prediction, the length scale of contraction strongly depends on myosin concentration. For low motor concentrations up to $R_M = 0.002$, the networks appear stationary for the entire duration of the experiment. Large-scale collective breathing fluctuations are visible, indicative of a strongly connected network, but the motors exert insufficient force to contract the network (Supplementary Movie 9). Increasing the motor concentration to $R_M = 0.005$ results in a drastic change: the entire network collapses into one large cluster mediated by a uniform global contraction (Fig. 4f,h; Supplementary Movie 10). However, a further increase of $R_M$ results in smaller clusters (Fig 4g,i; Supplementary Movie 11). At high motor densities, $\xi_1$ decreases in a manner consistent with the model's prediction (Fig 4j) and we again recover scale-free cluster size distributions (Supplementary Figure 3.)

These results lead to a counterintuitive consequence: in order to coordinate contractions over macroscopic length scales, *less* motor activity is needed. Increasing motor activity only yields small clusters.

**Motors can nucleate many concurrent ruptures**

In order to better understand the effect of force on cluster size, we consider the opposing limits of local and global contraction in our simulations. These two regimes are clearly separated by the critically connected regime, as evident in the schematic phase space diagram in Fig 5. The global contraction regime is located at the bottom-right corner, where motor forces are low and network connectivity is high. In this limit, networks are rigid, filaments remain straight, and the network deforms affinely[22]. On the opposite corner of the phase diagram, where connectivity is low and force is high, we find the local contraction regime. Such weakly connected, loose networks deform nonaffinely, and filaments are significantly bent.

We can interpret these limits by considering two relevant timescales, $\tau_{off}$ and $\tau_{relax}$. The first timescale is the characteristic crosslink unbinding time $\tau_{off} = k_{off}^{-1}$. The tension $T$ experienced by a crosslink depends on both the motor force $f$ and the network configuration, which can change over time. Although the full dependence of crosslink tension on motor force is complex, the qualitative behavior is clear: when filaments are straight, motor stress does not greatly induce crosslink tension; when filaments are bent, crosslinks experience tension (Supplementary Figure 4).

The second timescale, $\tau_{relax}$, is the time it takes for filaments in the network to relax in response to a crosslink unbinding event. We estimate the values of $\tau_{off}$ and $\tau_{relax}$ from previous work[31,38]:

$$\tau_{off,0} \sim 1\text{–}10\text{s} \qquad \tau_{relax} \sim 0.1\text{–}1\text{s}.$$

The above value for $\tau_{relax}$ is set by the thermal equilibration of individual filaments. It acts as an upper bound: forces can cause faster relaxation. Therefore in the absence of tension, $\tau_{off} > \tau_{relax}$.

We now consider how these timescales respond to the two limits of local and global contraction. In the global contraction limit, $f$ and $T$ are small, and $\tau_{off} > \tau_{relax}$ holds: once a crosslink unbinds, the network fully relaxes before the next crosslink unbinds. This well-known limit corresponds to a quasistatic process[39]. Boundary conditions determine how the network evolves in this limit: networks fixed at rigid boundaries build up stress and rupture via the nucleation of a large crack at a microscopic flaw, reminiscent of Griffith's criterion[40]. Unanchored networks contract affinely, or drive shape changes when coupled to deformable boundaries[41].

In the opposite limit of local contraction, $f$ and $T$ are large, and the network satisfies $\tau_{\text{off}} < \tau_{\text{relax}}$: strong internal driving causes crosslinks to unbind quickly. Many cracks that rupture the network into clusters form across the whole network, rather than nucleating at a single flaw. The presence of a finite viscosity in our model is essential for this behavior. Neglecting viscosity leads to $\tau_{\text{relax}} = 0$, and networks fail only via quasistatic crack propagation[39].

In between the two limits of global and local contractions, we find critically connected networks with a scale-free distribution of clusters. For zero force, this regime is narrow and centered around the critical point. As forces increase, this regime broadens and shifts to higher connectivities. This rightward shift reflects an asymmetry where motor activity reduces connectivity, rather than increasing it. The broadening shows that increased motor activity drives networks more robustly to a critical state.

Intriguingly, robust critical behavior has been demonstrated in many biological systems[42-46]. Internal driving could underlie robust criticality[47], but so could other mechanisms, including natural selection[48,49]. Disentangling these mechanisms cannot be addressed by studying living systems alone. Here we report robust criticality in a minimal model system and show that internal driving is directly responsible. These results may help explain criticality in other biological contexts and may prove useful in designing the physical properties of synthetic active materials, which have recently become available[50].

Our framework offers a minimal microscopic mechanism that should help in modeling contractile systems in biology. Recent studies in live cells suggest that motor myosin-driven cytoskeletal ruptures play an important functional role in cell division,[51] whereas they contribute to developmental defects in developing embryos[52]. Consistent with our findings, decreased connectivity caused dramatic rupture of the ventral furrow into clusters of cells in developing fly embryos. We anticipate that our framework applies more generally to tissues of interconnected cells[53,54], where a supracellular actomyosin network transmits forces over tissue length scales.

## Methods

**Protein Preparation**. Actin and myosin were prepared from rabbit psoas skeletal muscle (Supplementary Information). Myosin II was labeled with Alexa Fluor 488 NHS ester (Invitrogen, Paisley, UK); actin was labeled with Alexa Fluor 594 carboxylic acid, succinimidyl ester[13]. Recombinant mouse fascin was prepared from T7 pGEX *E. coli* [55].

**Sample Preparation**. Samples were mixed to yield a final buffer composition of 20 mM imidazole pH 7.4, 50 mM potassium chloride, 2 mM magnesium chloride, 1 mM dithiothreitol, and 0.1 mM adenosine triphosphate (ATP). Furthermore, 1 mM trolox, 2 mM protocatechuic acid, and 0.1 µM protocatechuate 3,4-dioxygenase were added to minimize photobleaching. The ATP level was held constant by addition of 10 mM creatine phosphate disodium and 0.1 mg mL$^{-1}$ creatine kinase. The actin concentration was held constant at 12 µM (0.5 mg mL$^{-1}$). Freshly mixed actoymyosin solutions were loaded onto polyethylene-glycol-passivated flowcells with a geometry of 2.5 x 2.5 x 0.1-mm$^3$ (Supplementary Information) and sealed with either Baysilone silicone grease (Bayer, Leverkusen, Germany) or uncured PDMS (Dow Chemicals, Midland, MI, USA). The time evolution of the network structure was observed with a Nikon PlanFluor 4x objective (NA 0.13), which allows the network to fit entirely within the objective's field of view.

**Image Analysis**. Cluster sizes were determined by a customized algorithm, implemented in MATLAB. Time-lapse images of contracting actomyosin networks were analyzed, starting from the final acquired frame. Cluster evolution, determined from Voronoi diagrams of myosin foci, was tracked by looping the algorithm backwards in time (Supplementary Information).

**Definition of $\xi_1$ and $\xi_2$**. For experimental results, we measure the areas $a_i$ of the initial network that contract together, which we define as clusters. We define $\xi_1$ as the weighted mean of cluster sizes $l_i$ (square root of area), in analogy to the definition of the correlation length from percolation theory[26]:

$$\xi_1 := \sum_i l_i a_i^2 / \sum_i a_i^2$$

This length scale is dominated by the largest cluster. We furthermore define $\xi_2$ in analogy to percolation theory:

$$\xi_2 := {\sum}'_i l_i a_i^2 / {\sum}'_i a_i^2$$

where ${\sum}'_i$ denotes summation over all clusters except for the largest cluster, as well as long edge clusters (Supplementary Information). This length scale is dominated by the second-largest cluster.

For simulation results, $\xi_1$ and $\xi_2$ are given by the square root of the harmonic-averaged area of the largest and second largest clusters, respectively, over 10-100 disorder realizations for each set of parameters.

## Acknowledgements

This work is part of the research programme of the Foundation for Fundamental Research on Matter (FOM), which is part of the Netherlands Organisation for Scientific Research (NWO). GK and JA were funded by a Vidi grant from the Netherlands Organization for Scientific Research (NWO). We thank M. Kuit-Vinkenoog, M. Preciado-López, and F.C. Tsai (AMOLF, Amsterdam, Netherlands) for help with purifications, S. Hansen and R.D. Mullins (UCSF, San Francisco, USA) for the fascin plasmid, K. Miura (EMBL, Heidelberg, Germany) for the Temporal Color Code ImageJ plugin, and C. Broedersz (Princeton University, NJ, USA) for insightful discussions.

## Author Contributions

J.A. and G.K. designed experiments. J.A. performed experiments. M.S., A.S., and F.C.M. designed simulations. M.S. and A.S. performed simulations. All authors contributed to the writing of the paper.

## Additional Information

The authors declare no competing financial interests.

**Figure Captions**

**Figure 1**: Experiments with motor-driven networks show that initial connectivity controls the length scale of contraction. **a**: Schematic representation of the experiment. Actin filaments (black lines) are connected by crosslinks (purple circles), and myosin motors (green dumbbells) exert force dipoles (orange arrows) on actin filaments. **b–d**: Temporal evolution of three networks with varying amounts of fascin crosslinks (a. $R_C = 0.01$; b. $R_C = 0.05$; c. $R_C = 0.1$). Actin and motor concentrations are constant: [actin] = 12 µM; $R_M = 0.01$. Color corresponds to time according to calibration bar (b, left). Times ($t_{start}$, $t_{end}$) in minutes after initiation of actin polymerization: b. (2, 20); c. (2,120); d. (1,5). Scale bar 1 mm. See Supplementary Movies 1–3. **e–g**: Decomposition into clusters, delimited by black lines. Color indicates the largest (blue) and the second-largest (pink) cluster, whose sizes correspond to $\xi_1$ and $\xi_2$ respectively. Note that (g) does not have a second-largest cluster because we exclude long edge domains from our analysis (Supplementary Figure 6). **h**: Dependence of $\xi_1$ (blue circles) and $\xi_2$ (pink triangles) on crosslink concentration ($R_C$). Error bars denote standard errors of the mean for repeat experiments: 1, 6, 13, 14, 9, and 5 experiments for $R_C = 0.002, 0.005, 0.01, 0.02, 0.05$, and $0.1$, respectively. *Inset*: Predicted dependence of $\xi_1$ and $\xi_2$ on connection probability $p$ according to percolation theory, given experimental parameters (Supplementary Information).

**Figure 2**: Cluster size distributions depend on network connectivity, exhibiting power-law distributions when $\xi_1 \sim \xi_2 \sim L$. **a**. Scatter plot of 48 samples with different $R_C$ in $\xi_1$-$\xi_2$-space (see legend, top left). Boxes delimit different regimes: local contraction ($\xi_1 < 300$ µm), critically connected ($\xi_1 \geq 300$ µm and $\xi_2 \geq 300$ µm), and global contraction ($\xi_1 \geq 1500$ µm and $\xi_2 < 300$

µm). Two data points with $\xi_2 = 0$ are depicted here with $\xi_2 = 30$ µm. **b**: Histogram (circles) and complementary cumulative probability distribution (solid lines) of cluster areas, $a$ / µm², for the three regimes. For the critically connected regime, data across more than two orders of magnitude (red circles) are statistically consistent with a power-law distribution (solid red lines) with an exponent of –1.91 ± 0.06, $p = 0.52$, where $p > 0.1$ indicates plausible agreement with a power law (Supplementary Information). Note that the slope of the complementary cumulative probability distribution is equal to one plus the slope of the histogram because the histogram is the absolute value of the derivative of the complementary cumulative probability distribution.

**Figure 3**: Simulations show that motors can drive initially well-connected networks to a critical state. **a**: Schematic representation of the simulation. A triangular lattice of nodes, connected by line segments (black lines), contains an average of $N$ crosslinks per node (purple circles). During the course of the simulation, pairs of nodes experience contractile forces (orange arrows) and move in response to these forces. **b**: Temporal evolution of a representative network in the absence of remodeling. **c**: Motors cause network restructuring by generating tension $T$ on crosslinks that increases the off-rate $k_{off}$. **d-j**: Simulated networks exhibit behavior consistent with experiment. See Supplementary Movies 4–6. **d,f,g**: Temporal evolution of three networks differing in initial connectivity: a. $c = 0.025$; b. $c = 3$; c. $c = 10000$. Force is constant: $f / k = 50$. Color corresponds to simulation time according to calibration bar (d, left). Box size $L$ is 100 times longer than the initial lattice size $l_0$. **e,g,i**: Decomposition into clusters, shaded by pastel colors. Bold color indicates the largest (blue) and the second-largest (pink) clusters, whose sizes correspond to $\xi_1$ and $\xi_2$ respectively. **j**: Dependence of $\xi_1$ (blue circles) and $\xi_2$ (pink triangles) on crosslink concentration $c$ across repeat simulations. Open symbols indicate values at $t = 0$, which corresponds to passive networks described by classical percolation theory. Closed symbols

indicate values at the end of the simulation, after the network has broken up into clusters. Yellow regions correspond to values of *c* for which $\xi_2 > L / 10$ and the cluster size distribution exhibits a power-law. Note that this region is narrow for classical percolation theory (diagonal yellow stripes) but broadens substantially in response to active internal driving (solid yellow box).

**Figure 4**: Simulation and experiment both show that increased motor force reduces cluster size. **a,b**: Temporal evolution of two simulated networks with constant network connectivity (*c* = 3) but with either (a) low force, $f / k = 3$; or (b) high force, $f / k = 150$. Color corresponds to simulation time. **c,d**: Decomposition into clusters. **e**: Dependence of $\xi_1$ (blue circles) and $\xi_2$ (pink triangles) on force *f*. **f,g**: Temporal evolution of two experimentally prepared networks with (f) low myosin concentration, $R_M = 0.005$; or (g) high myosin concentration, $R_M = 0.02$. Color corresponds to time. Times ($t_{start}$, $t_{end}$) in minutes after initiation of actin polymerization: a. (2, 43); b. (2,14). The network connectivity is the same in both cases ([actin] = 12 µM, $R_C = 0.02$). See Supplementary Movies 8–10. **h,i**: Decomposition into clusters. **j**: Dependence of $\xi_1$ (blue circles) and $\xi_2$ (pink triangles) on myosin concentration, given by $R_M$. Scale bars 1 mm. Error bars denote standard errors of the mean for repeat experiments: 5, 14, and 5 experiments for $R_M$ = 0.005, 0.01, and 0.02, respectively. Dashed lines depict $f^{-1}$ (panel e) and $R_M^{-1}$ (panel j).

**Figure 5**: The critically connected regime broadens with increasing force. **a**: Schematic of proposed phase diagram in force-connectivity space, where the critically connected regime separates the local contraction and global contraction regimes. **b, c**: Dependence of $\xi_1$ (b) and $\xi_2$ (c) simulated over a broad range of force and connectivity.

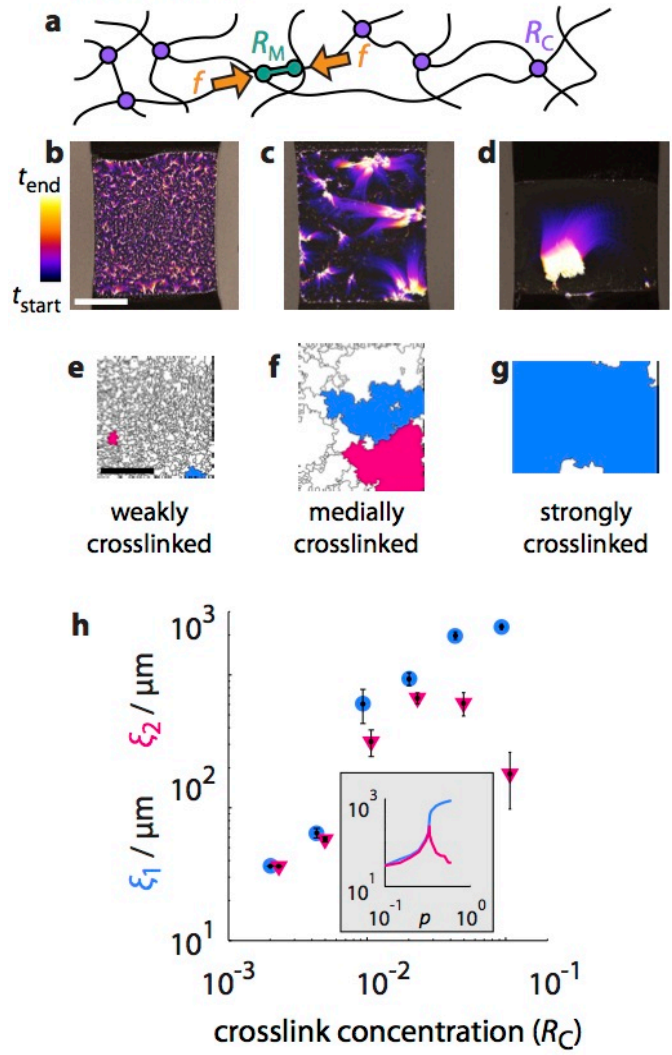

**Experiment**

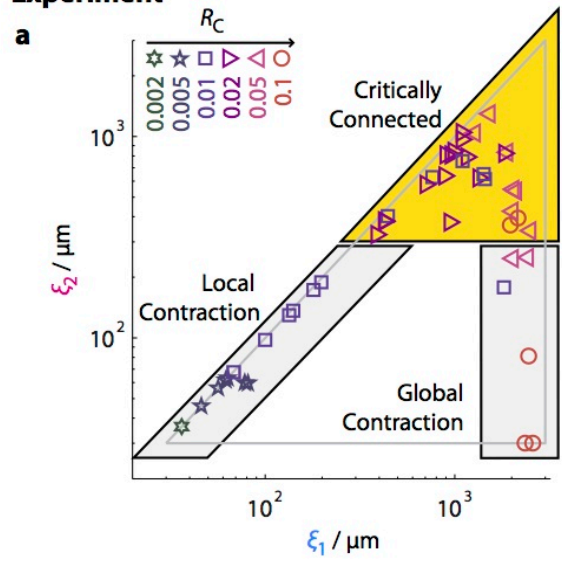

a

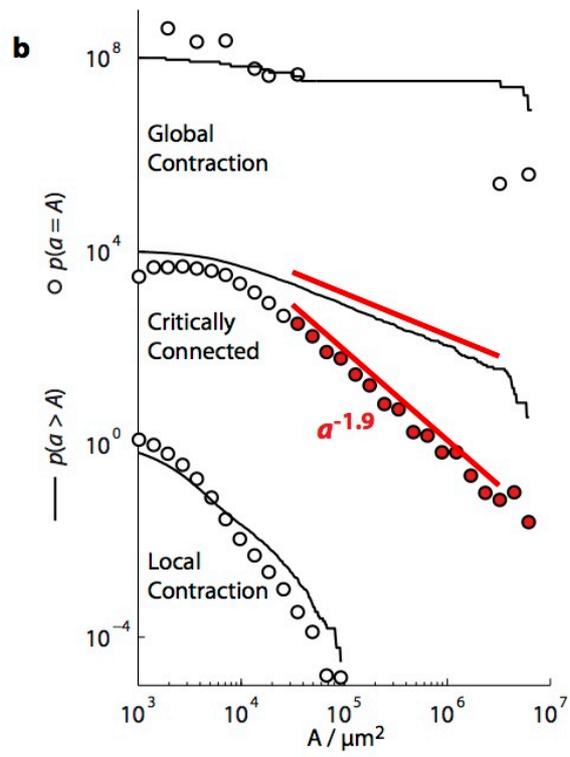

b

**Simulation**

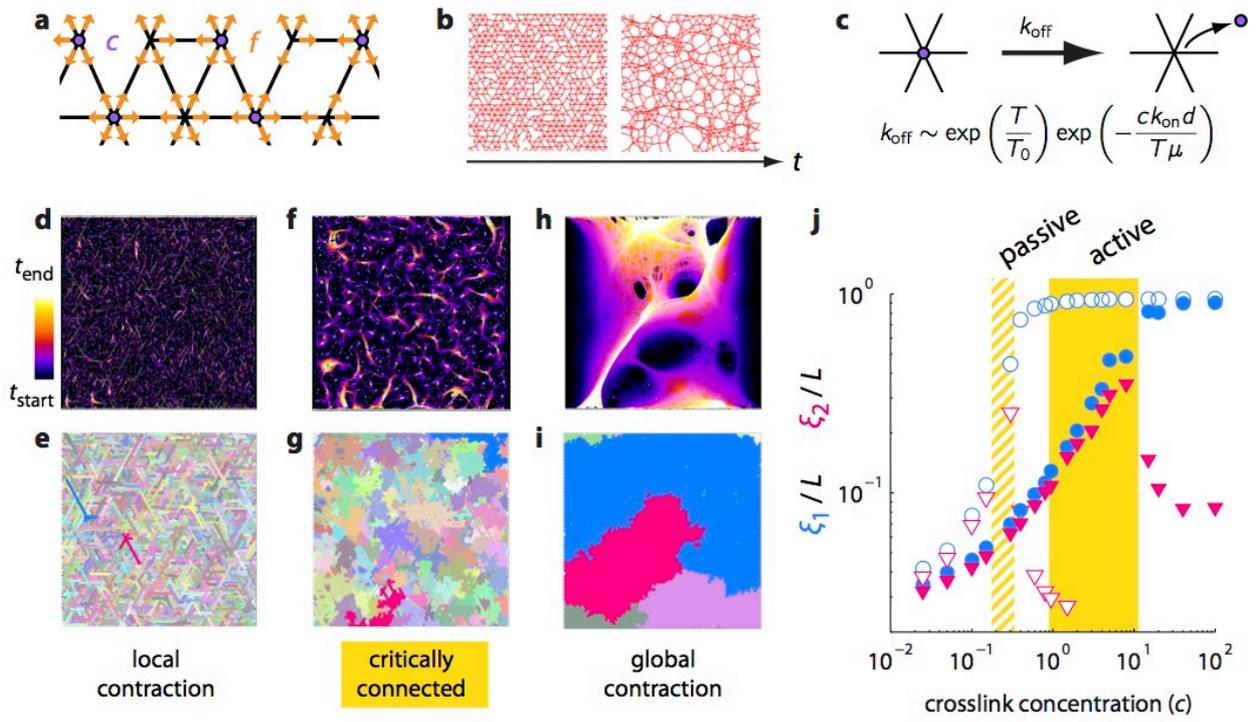

local contraction | critically connected | global contraction

**Simulation**

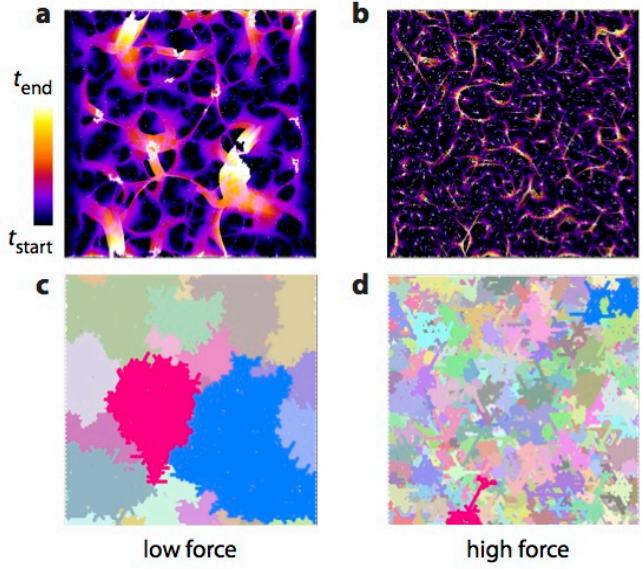
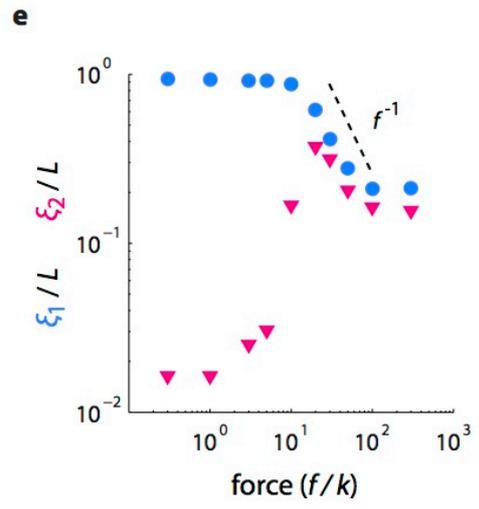

low force     high force

**Experiment**

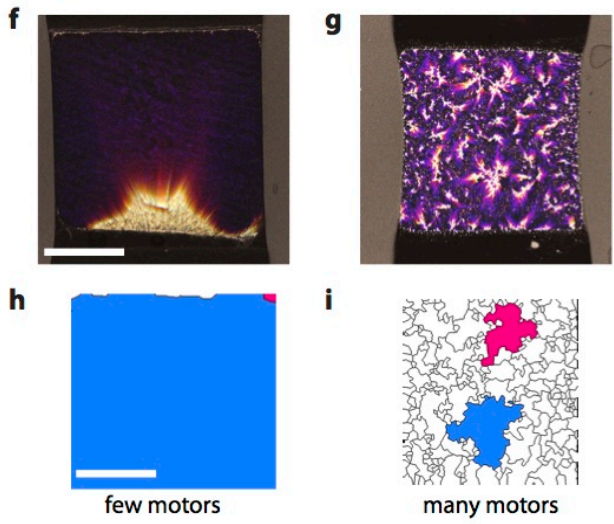
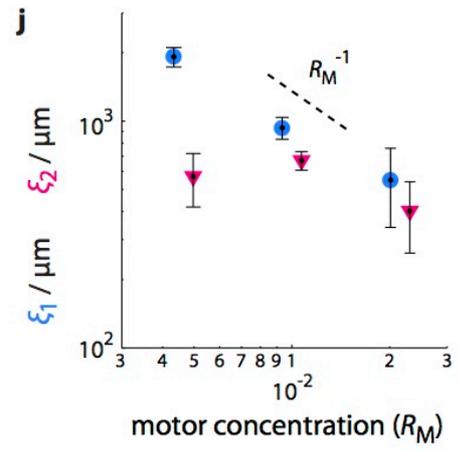

few motors     many motors

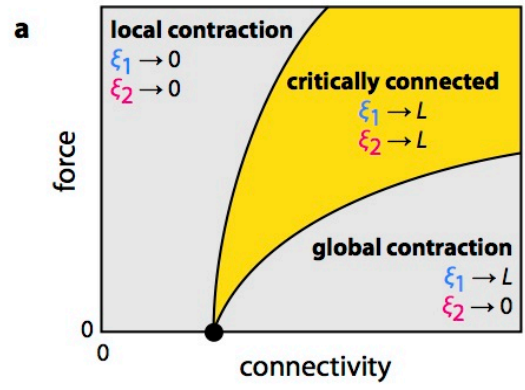

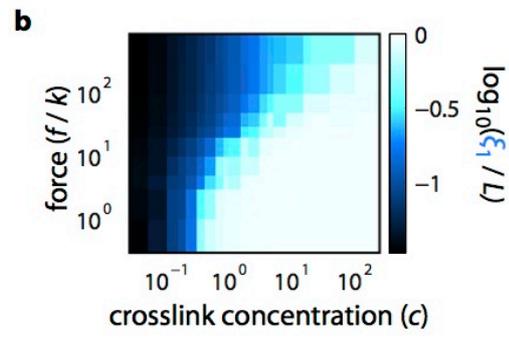

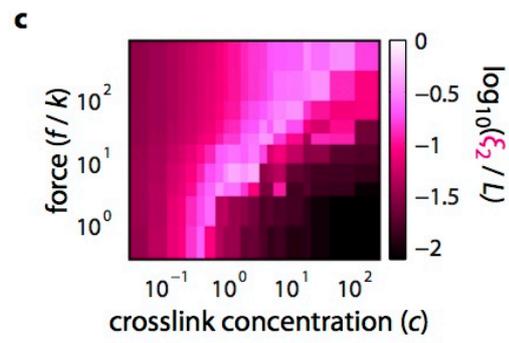

**Molecular motors robustly drive active gels to a critically connected state**

José Alvarado[1], Misha Sheinman[2], Abhinav Sharma[2], Fred MacKintosh[2], Gijsje Koenderink[1]

S U P P L E M E N T A R Y   I N F O R M A T I O N

**Methods**

**Protein preparation.** Monomeric (G-) actin and myosin II were purified from rabbit psoas skeletal muscle[1]. G-actin was purified with a Superdex 200 column (GE Healthcare, Waukesha, WI, USA) and stored at −80 °C in G-buffer (2 mM tris-hydrochloride pH 8.0, 0.2 mM disodium adenosine triphosphate, 0.2 mM calcium chloride, 0.2 mM dithiothreitol). Myosin II was stored at −20 °C in a high-salt storage buffer with glycerol (25 mM monopotassium phosphate pH 6.5, 600 mM potassium chloride, 10 mM ethylenediaminetetraacetic acid, 1 mM dithiothreitol, 50% w/w glycerol). Creatine phosphate disodium and creatine kinase were purchased from Roche Diagnostics (Indianapolis, IN, USA), all other chemicals from Sigma Aldrich (St. Louis, MO, USA). Magnesium adenosine triphosphate was prepared as a 100 mM stock solution using equimolar amounts of disodium adenosine triphosphate and magnesium chloride in 10 mM imidazole pH 7.4.

**Sample Preparation.** Fresh myosin solutions were prepared by overnight dialysis into myosin buffer (20 mM imidazole pH 7.4, 300 mM potassium chloride, 4 mM magnesium

[1] FOM Institute AMOLF, Science Park 104, 1098 XG Amsterdam, The Netherlands
[2] Department of Physics and Astronomy, Vrije Universiteit, 1081 HV Amsterdam, The Netherlands

chloride, 1 mM dithiothreitol) and used within four days. All frozen protein stocks (actin, myosin, fascin) were clarified of aggregated proteins upon thawing at 120,000 $g$ for at least 5 min and used within four days. The proteins' concentrations in the supernatant were determined by measuring the solution absorbance at 280 nm with a NanoDrop 2000 (ThermoScientific, Wilmington, DE, USA) and using extinction coefficients, in $M^{-1}$ $cm^{-1}$, of 26600 (actin[2]), 249000 (myosin[3]) and 66280 (fascin, computed from amino acid sequence[4]). Fluorescently labeled proteins were mixed with unlabeled proteins to yield a 10% molar ratio of dye to protein. During sample preparation, myosin and Alexa-488-myosin were mixed at high salt and then mixed into a tube containing fascin and buffer. This solution was mixed into a second tube containing actin and Alexa-594-actin to initiate polymerization and immediately inserted into glass flowcells passivated by poly-L-lysine-polyethylene-glycol (Surface Solutions AG, Dübendorf, Switzerland).

**Preparation of flow cells**. Glass flow cells were assembled by sandwiching strips of ParaFilm between a long cover slip (24 mm x 60 mm) and 2.5-mm-narrow glass strips which were manually cut from 40-mm-long cover slips. This yielded 2.5 x 2.5 x 0.1-$mm^3$-large chambers (corresponding to ~0.6 µL). All glass was cleaned with piranha solution, rinsed in MilliQ water, and stored in isopropanol. Assembled flow cells were then passivated by applying 1M potassium hydroxide for 5 min, rinsing with MilliQ, drying with $N_2$, applying 0.2 mg $mL^{-1}$ poly-L-lysine-polyethylene-glycol (Surface Solutions AG, Dübendorf, Switzerland) for 30 min, rinsing with MilliQ, and drying with $N_2$.

**Simulation**. The values taken for the simulations are: the system size $W=100$, $k_{off,0}=10$, $T_0=1$, $k_{on}\ d/\ \mu=10$, $k=1$. The buckling is implemented by vanishing force of a bond for a compression strain below 0.1. The stiffening is implemented by increase of the stretching constant by 100-fold for extension strain above 0.2.

## Algorithm for determining cluster size

We developed a MATLAB algorithm to determine sizes of contracting clusters from time-lapse images of contractile actomyosin networks. Actin filaments and myosin motors were fluorescently labeled to appear in separate channels. In short, this technique begins with the final frame of acquisition (Fig 5a), determines clusters, and tracks the expansion of these clusters back in time until the first frame of acquisition. The result is a decomposition of the initial network into clusters.

**Step 1**: the final acquired image from the actin channel (Fig 5b) was median-filtered with a radius of 1px (Fig 5c). This step filters out noise.

**Step 2**: the median-filtered image was thresholded using Otsu's method[5] (Fig 5d). The result of this step is a binary image of only black or white pixels. Contiguous groups of white pixels are called connected components. Each connected component corresponds to a cluster of actin and myosin.

**Step 3**: the final acquired image from the myosin channel (Fig 5e) was also median filtered with a radius of 1px (Fig 5f).

**Step 4**: the median-filtered image was thresholded using Otsu's method (Fig 5g), again yielding a black-and-white image of connected components that correspond to clusters of actin and myosin.

**Step 5**: the thresholded image was morphologically opened (successive dilation and erosion) using a 1-px-radius-disk as a structuring element (Fig 5h). This step serves as an additional filter, removing connected components smaller than the structuring element.

**Step 6**: connected components from step 5 were assigned to connected components from step 2 (Fig 5i). Note that the connected components from step 2 (actin) are usually large, and contain many smaller connected components from step 5 (myosin). Without this step, the disjoint connected components from step 5 could be erroneously treated as separate clusters.

**Step 7**: domains were defined around each cluster, using the MATLAB function bwdist, which performs a distance transform (Fig 5j). This step decomposes the entire image into a Voronoi-like diagram, where domain boundaries occur halfway between connected components.

**Step 8**: steps 3-7 were repeated for the myosin channel, starting with the final acquired frame (Fig 5k), looping through successive acquired images backwards in time (Fig 5l), and finally arriving at the first acquired frame (Fig 5m). The end result is the first acquired frame, where the actin and myosin signals are uniformly distributed, and decomposed into clusters. During the loop, steps 3-7 were unchanged, except for step 6: myosin connected components were joined not by using actin connected components, but by the domains from the previous iteration of step 7.

**Step 9**: finally, the image of domains produced from step 7 of the final loop was cropped to the largest rectangle contained by the network.

In two cases, adjustments to this routine were necessary. In one case, during step 2, Otsu's threshold sometimes yielded large connected components that spanned the image (Fig 6a). This resulted in the network being erroneously represented as one large cluster (Fig 6b). This artifact was eliminated by choosing a more restrictive threshold (Fig 6c), which resulted in accurate domains (compare panels *d* and *e*). In another case, during step 5, morphological opening sometimes filtered out small, dim clusters (Fig 6f). This led to their corresponding domains to disappear, and small, neighboring clusters were reported as bigger clusters (Fig 6g). This artifact was eliminated by omitting step 5 (Fig 6h), yielding accurate domains (compare panels *i* and *j*).

## Adjustment of domains

We performed two types of adjustments to the cluster decompositions produced by our algorithm. First, we removed long edge domains from our analysis. These domains could be the result of enhanced interactions with the edge of the confining geometry, in addition to internal driving by myosin activity. We first search for domains that touch the border of the cropping rectangle (Fig 5m, dashed line). We next compute the major and minor axis of the ellipse that has the same normalized second central moment of each domain, as well as the orientation of the major axis. Finally, we consider edge domains which satisfy the following two conditions: (i) the major-axis-to-minor-axis ratio is greater than two; (ii) the major axis is oriented along the edge that the domain touches to within 45º. (Condition ii is dispensed for corner domains that touch two edges.) Edge domains that satisfy these two conditions are then omitted from the sums in the definition of $\xi_1$ and $\xi_2$, as well as when plotting distributions.

Second, we compensated for fast clusters. Sometimes the displacement of a cluster between two successive frames was greater than the half-way distance to a neighboring cluster. Clusters would then leave their own domain and erroneously enter neighboring domains. This artifact mostly affected networks in the global contraction regime, where the global build-up of stress led to fast relaxation events. This artifact cannot be addressed by modifying the algorithm. We therefore manually corrected contraction domains to accurately reflect network evolution. A total of five corrections were performed, all of which are reported in Fig 7.

These two adjustments to domain size affect our results for the global contraction regime. This is evident by inspecting the effect of the two adjustments on cluster size distributions (Fig 8a). However, the local contraction and critically connected regimes are largely unaffected. The power-law exponents determined from experiment are robust to the two adjustments described above (Fig 8b).

**Statistical analysis of domain sizes**

In order to determine whether cluster size distributions were consistent with a power law, we employed a recently developed, rigorous statistical analysis[6]. This technique first fits observed data to a power law, determining both the best exponent and best lower cut-off. The lower cut-off is the minimum value above which the power law is fitted. It then compares the observed dataset with multiple synthetic datasets (generated from a power-law distribution using the best fit parameters) by computing the Kolmogorov-Smirnov statistic, which quantifies the "distance" between a dataset and the true power-law distribution. Finally, it computes a *p*-value, which is defined as the fraction of synthetic datasets whose distance is greater than the observed dataset. Therefore, larger *p*-values correspond to an increased likelihood that the observed dataset is consistent with a power law. A power law can be ruled out for $p < 0.1$.

**Percolation model**

Our model is based on three-dimensional network of $N$ straight filaments of length $L$ placed in a $W$ x $W$ x $W$ box. The filaments are placed such that their position and orientation is uniformly distributed. Two filaments are considered to be intersecting if the shortest distance between them is less than a certain value which is taken to be of the order of size of the cross link. At this intersection these two filaments can be connected by a freely hinging crosslink. The probability that such a crosslink exists is denoted by *p*. Periodic boundary conditions are assumed in all directions. The line density $NL/W^3$ is obtained from the experiments, and estimated to be ~20 $\mu m^{-2}$. Our simulations show that the connectivity percolation occurs in the vicinity of *p*=0.33.

**Figure Captions**

**Figure 1**: Confocal image of actin (red) and myosin (cyan) in the absence of crosslinks ([actin] = 12 μM, $R_M$ = 0.01, $t \sim 2$ h after the initiation of actin polymerization). Myosin motors form small foci, which are separated approximately 30 μm apart.

**Figure 2**: Simulated histogram (open circles) and corresponding cumulative probability distribution (closed circles) of cluster areas, $a / l_0^2$, for the three conditions shown in the main text, Fig 3d-f. Solid red lines denote a power law with exponent −2.0.

**Figure 3**: Increased myosin activity results in smaller clusters. **a**: Scatter plot of samples with different $R_M$ in $\xi_1$-$\xi_2$-space (see legend, top left). Crosslink concentration is constant ($R_F = 0.02$). Boxes delimit different regimes: local contraction ($\xi_1 < 300$ μm), critically connected ($\xi_1 \geq 300$ μm and $\xi_2 \geq 300$ μm), and global contraction ($\xi_1 \geq 1500$ μm and $\xi_2 < 300$ μm). **b**: Histogram (circles) and corresponding cumulative probability distribution (solid lines) of cluster areas, $a /$ μm$^2$, for the three regimes. For the critically connected regime, data across more than two orders of magnitude are consistent with a power-law distribution with an exponent of $-1.90 \pm 0.06$ ($p = 0.12$, $a_{min} = (20 \pm 8) \, 10^3$ μm$^2$).

**Figure 4**: Bent filaments induce tension on crosslinks. **a**: Two straight filaments (black lines) are crosslinked (purple circle) at their intersection. If forces (orange arrows) are balanced, the crosslink experiences zero tension. This is evident because if the crosslink unbinds (right), no relaxation occurs. **b**: A straight filament and a bent filament are crosslinked. Although forces are balanced, the crosslink here experiences tension. This is evident because if the crosslink unbinds (right), the bent filament relaxes to a straight conformation.

**Figure 5**. Cluster-size algorithm. See Supplementary Movie 11. **a**: Final image of a time-lapse acquisition of a contractile actomyosin network ($t = 80$ min). Alexa-594-actin is shown in red, DyLight-488-myosin in green. [actin] = 12 µM, $R_F = 0.02$, $R_M = 0.02$. **b**: Close-up of the actin channel, corresponding to dashed box, *a*. **c**: Median filter of *b*. **d**: Otsu threshold of *c*. **e**: Close-up of the myosin channel, corresponding to dashed box, *a*. **f**: Median threshold of *e*. **g**: Otsu threshold of *f*. **h**: Morphological opening of *g*. **i**: Superposition of *d* (red) and *h* (green). Note that some disjoint green clusters are contained within one large red cluster. In this case, they are treated as one large cluster. **j**: Superposition of original myosin signal (white) with domains (shades of beige), which result from a distance transform, implemented in MATLAB as the *bwdist* function. **k**: Myosin signal (white) and domains (shades of beige) for the entire sample before looping the algorithm, beginning with the final frame ($t = 80$ min). **l**: Close-up of *k* (dashed box) at three representative stages of loop progression as clusters expand in -*t* (from left to right: $t = 20$ min, 8 min, 4 min). **m**: Myosin signal and resulting clusters for the first acquired frame ($t = 2$ min). Note that myosin is distributed homogeneously across the entire network at the beginning of acquisition, and is fully decomposed into clusters. **n**: Clusters (outlined in black) after cropping to dashed box, *k*, with largest and second-largest clusters denoted in blue and pink, respectively. Scale bars: a: 1 mm, b-j: 200 µm, k: 1 mm, l: 500 µm, m: 1 mm.

**Figure 6**. Two modifications of the cluster-size algorithm. **a-e**: First modification: skipping step 2 when it erroneously yields large, system-spanning connected clusters. **a**: Actin channel of original image ($R_F = 0.01$, $R_M = 0.01$, local contraction regime). **b**: Result of step 2. Note that the thresholded image does not resemble the individual clusters visible in the original image. **c**: Result of continuing the algorithm, which erroneously represents the sample as one large cluster. **d**: Result of the algorithm, skipping steps 1 and 2. Note that this image correctly represents

individual clusters. **e**: Overlay of acquired data, where color corresponds to time (calibration bar, top right). $t_{start}$ = 1 min; $t_{end}$ = 20 min. Note that this image qualitatively captures cluster evolution, and is obtained independently of the cluster-size algorithm. Comparing to panels *c* and *d* shows that panel *d* more accurately represents true cluster size. **f-l**: Second modification: skipping step 5 when it removes small, dim myosin clusters. **f**: Result of step 4 ($R_F$ = 0.01, $R_M$ = 0.01, local contraction regime). **g**: Close-up of *f*, green dashed box. **h**: Result of step 5. **i**: Close-up of *h*, green dashed box. Note that morphological opening removes very small clusters. **j**: Result of continuing the algorithm, which erroneously joins together many small clusters in one large cluster. **k**: Result of the algorithm, skipping step 5. Note that this image correctly represents individual clusters. **l**: Time overlay, as in *e*. $t_{start}$ = 2 min; $t_{end}$ = 30 min. Comparing to panels *j* and *k* shows that panel *k* more accurately represents true cluster size.

**Figure 7**: Manual correction of five experiments. True sample dynamics is depicted in the time overlay (first column). For these five experiments, the algorithm produces excessively large domains (second column). Upon careful visual inspection of the original data, erroneous domains were manually corrected to their apparent true size (third column). Scale bar 1 mm.

**Figure 8**: Results from modifying algorithm output. **a**: Distributions of cluster sizes ($R_M$ = 0.01) that result when either removing long edge domains (rows) or manually correcting domains (columns). Distributions are divided according to global contraction (top), critically connected (center), and local contraction (bottom) (see main text, Figure 2).

## Movie Captions

**Movie 1:** Experiment depicting the time-lapse of a weakly connected network ($R_F$ = 0.01, $R_M$ = 0.01). The initially homogeneous actomyosin network breaks up into small clusters. Actin is shown in red, and myosin is shown in cyan (white denotes overlap). Time after initiation of actin polymerization is shown at the top-right corner in hours:minutes:seconds.

**Movie 2**: Experiment depicting the time-lapse of a medially connected network ($R_F = 0.02$, $R_M = 0.01$). The initially homogeneous actomyosin network breaks up into large and small clusters. Actin is shown in red, and myosin is shown in cyan (white denotes overlap). Time after initiation of actin polymerization is shown at the top-right corner in hours:minutes:seconds.

**Movie 3**: Experiment depicting the time-lapse of a strongly connected network ($R_F = 0.01$, $R_M = 0.01$). The initially homogeneous actomyosin network contracts into one large cluster. A slower, secondary contraction subsequently occurs. Actin is shown in red, and myosin is shown in cyan (white denotes overlap). Time after initiation of actin polymerization is shown at the top-right corner in hours:minutes:seconds.

**Movie 4**: Custom MATLAB algorithm for determining cluster size. Sample depicted corresponds to Movie 11. White pixels correspond to myosin clusters. Beige regions correspond to domains around myosin clusters. The algorithm tracks cluster evolution backwards in time, yielding the original network, decomposed into clusters.

**Movie 5**: Experiment depicting a typical rupture event ($R_F = 0.02$, $R_M = 0.01$). Myosins exert forces that rupture the actin network, decreasing connectivity. Actin is shown in red, and myosin is shown in cyan (white denotes overlap). Time after initiation of actin polymerization is shown at the top-right corner in hours:minutes:seconds.

**Movie 6**: Simulation depicting the time-lapse of a locally contracting network ($c = 0.025$). The initially well-connected network breaks up into many small clusters. Nodes of a triangular lattice representing crosslink points (black dots) are connected by line segments representing actin filaments (red lines).

**Movie 7**: Simulation depicting the time-lapse of a critically connected network ($c = 3$). The initially well-connected network breaks up into small and large clusters. Nodes of a triangular lattice representing crosslink points (black dots) are connected by line segments representing actin filaments (red lines)

**Movie 8**: Simulation depicting the time-lapse of a locally contracting network ($c = 10000$). The initially well-connected network breaks primarily due to large crack. Nodes of a triangular lattice representing crosslink points (black dots) are connected by line segments representing actin filaments (red lines)

**Movie 9**: Experiment depicting the time-lapse of a network with very few motors ($R_F = 0.02$, $R_M = 0.002$). The actomyosin network does not contract. Rather, it exhibits slow rearrangements on macroscopic length scales, reminiscent of breathing, and indicative of a well-connected network. Actin is shown in red, and myosin is shown in cyan (white denotes overlap). Time after initiation of actin polymerization is shown at the top-right corner in hours:minutes:seconds.

**Movie 10**: Experiment depicting the time-lapse of a network with few motors ($R_F = 0.02$, $R_M = 0.005$). The actomyosin network contracts into one large cluster. Actin is shown in red, and myosin is shown in cyan (white denotes overlap). Time after initiation of actin polymerization is shown at the top-right corner in hours:minutes:seconds.

**Movie 11**: Experiment depicting the time-lapse of a network with many motors ($R_F$ = 0.02, $R_M$ = 0.02). The initially homogeneous actomyosin network breaks apart into clusters. Actin is shown in red, and myosin is shown in cyan (white denotes overlap). Time after initiation of actin polymerization is shown at the top-right corner in hours:minutes:seconds.

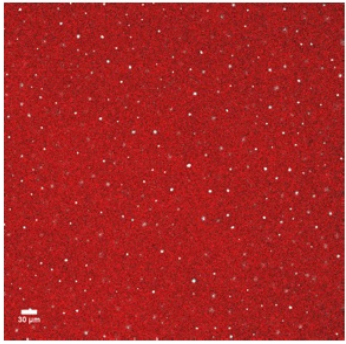

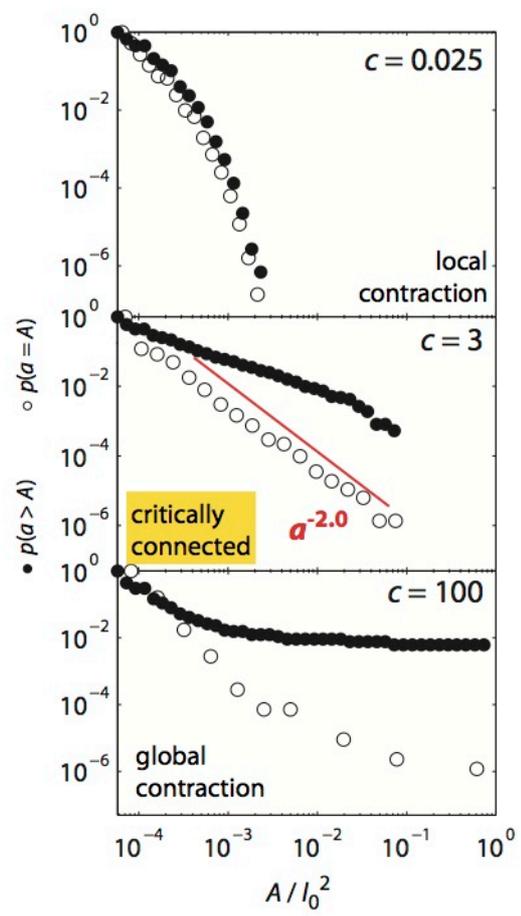

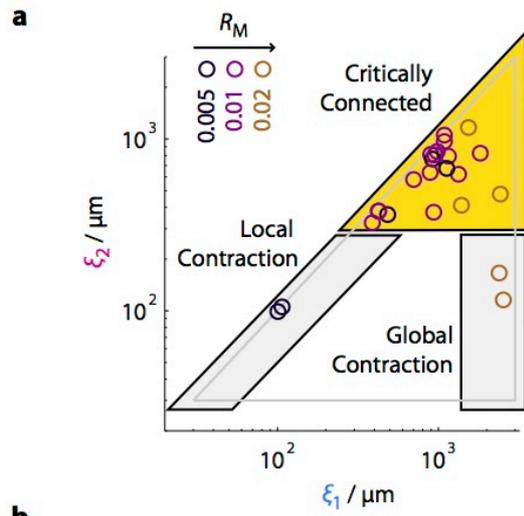
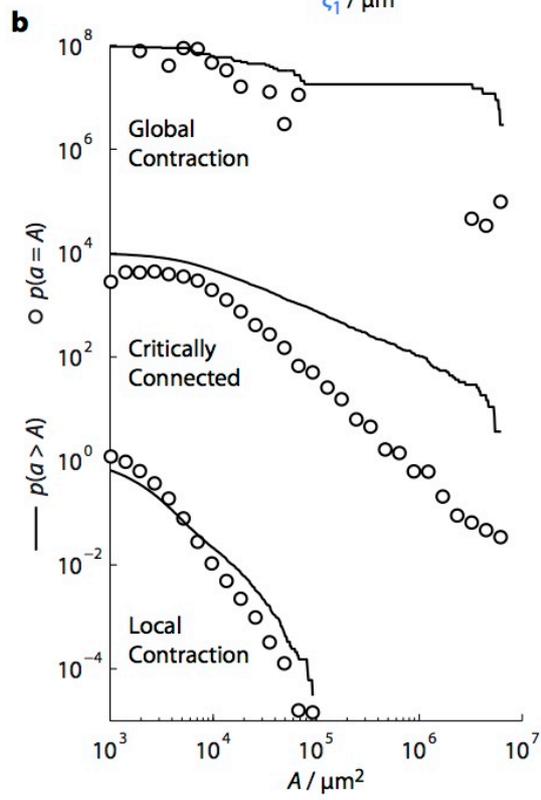

a

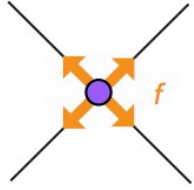 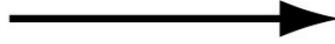 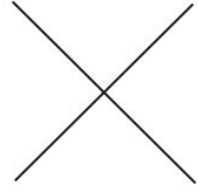

Filaments are straight     Crosslink unbinds     No relaxation: no tension stored

b

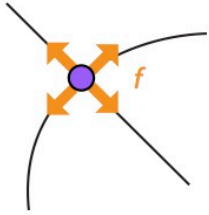 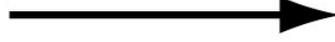 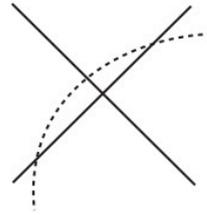

Filaments are bent     Crosslink unbinds     Bent filaments relax: crosslink stored tension

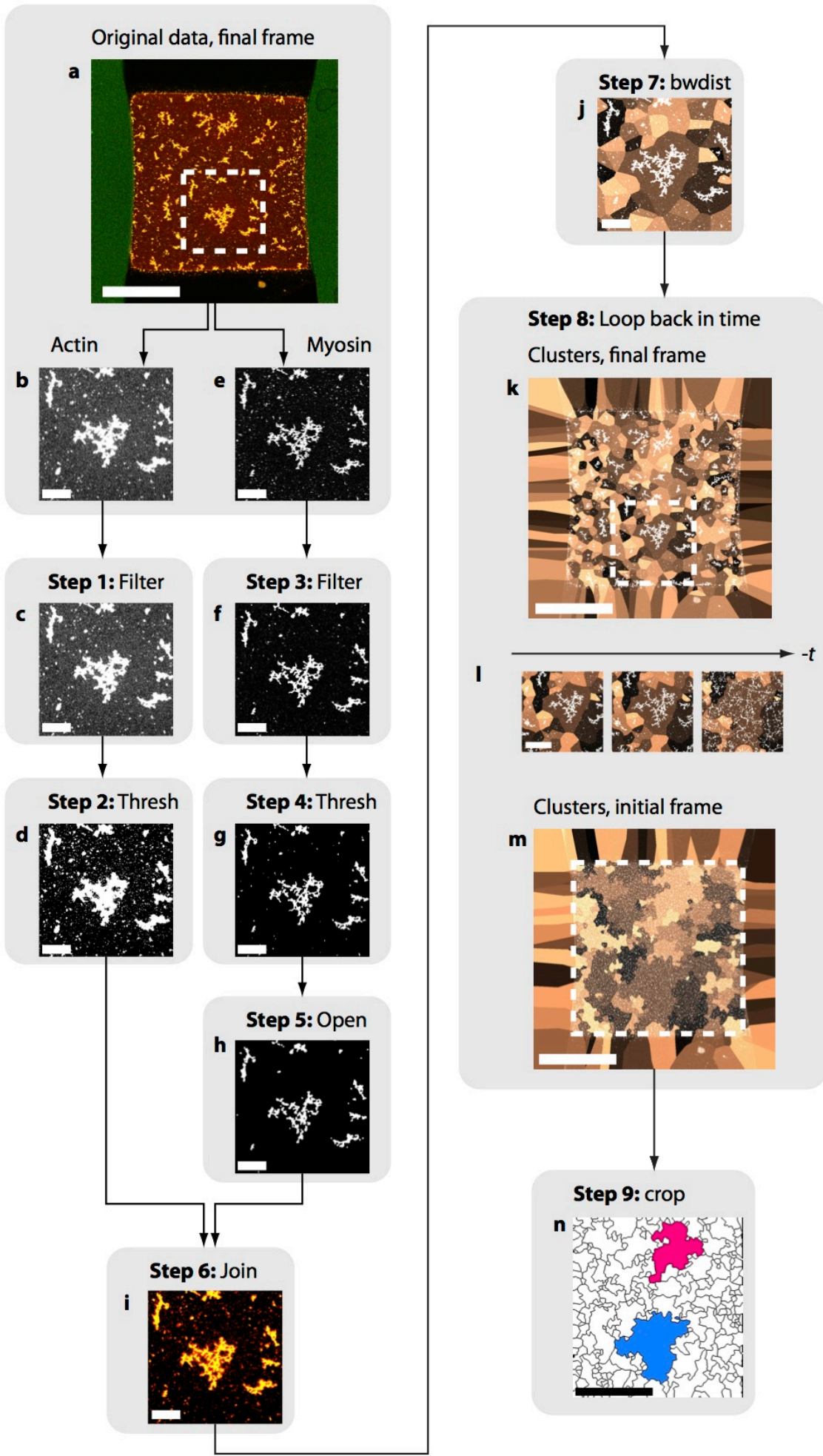

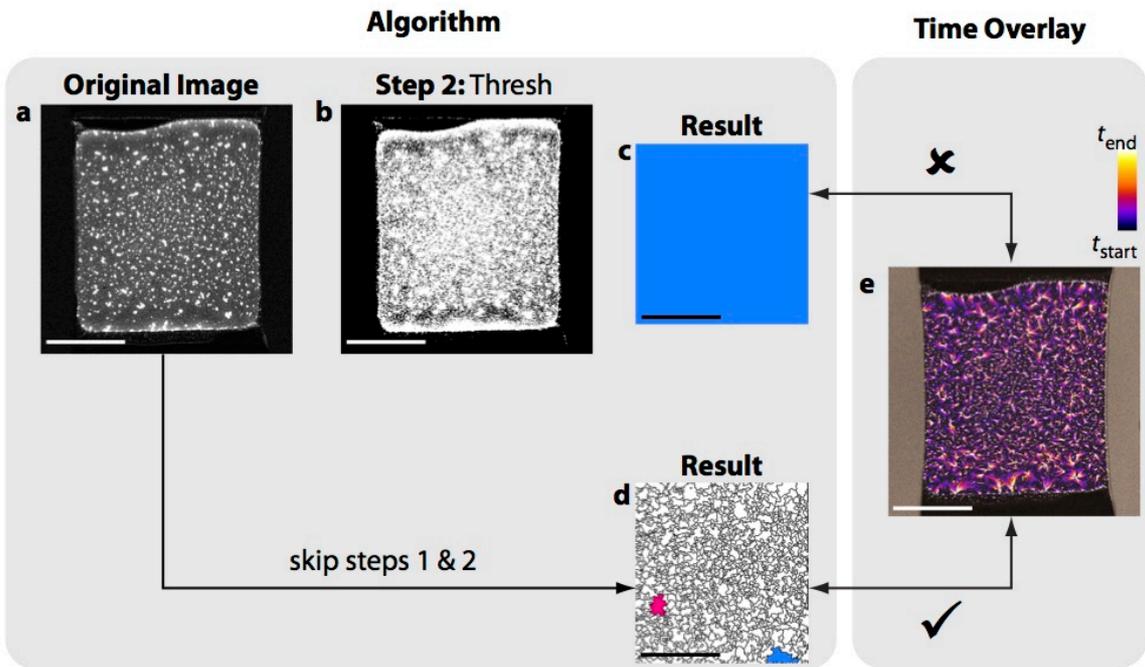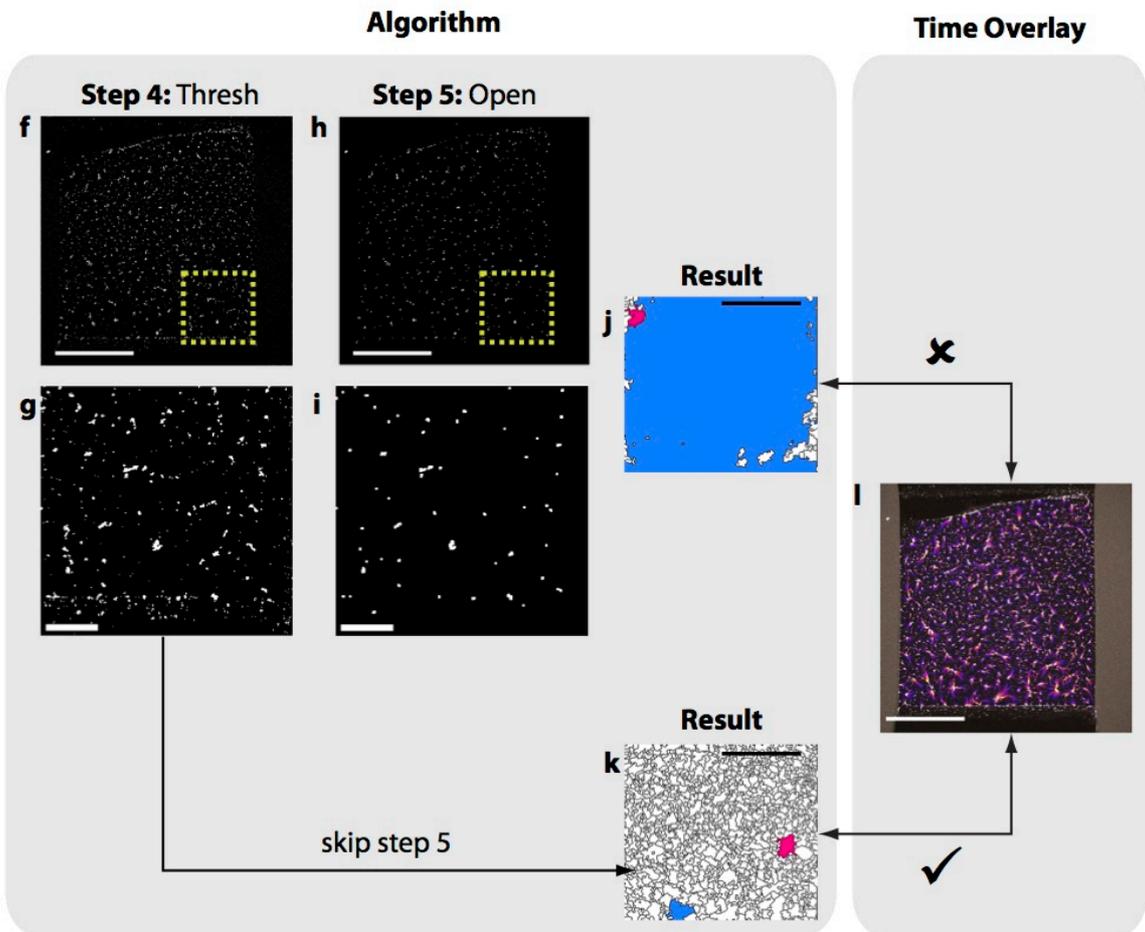

| Time Overlay | Result of Algorithm | Corrected |
|---|---|---|
| 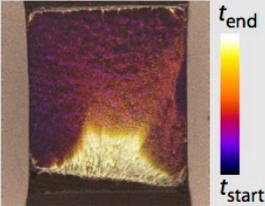 $R_F = 0.02$ $R_M = 0.005$ | 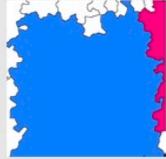 $\xi_1 = 2135$ μm $\xi_2 = 488$ μm | 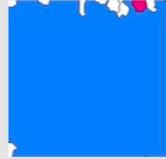 $\xi_1 = 2387$ μm $\xi_2 = 163$ μm |
| 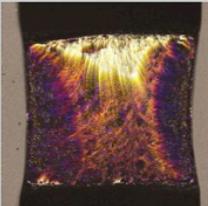 $R_F = 0.02$ $R_M = 0.005$ | 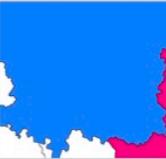 $\xi_1 = 2281$ μm $\xi_2 = 580$ μm | 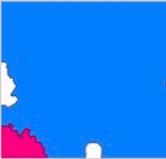 $\xi_1 = 2435$ μm $\xi_2 = 456$ μm |
| 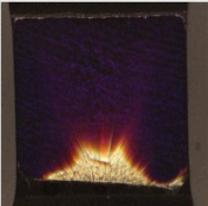 $R_F = 0.02$ $R_M = 0.005$ | 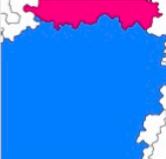 $\xi_1 = 2205$ μm $\xi_2 = 716$ μm | 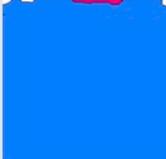 $\xi_1 = 2535$ μm $\xi_2 = 141$ μm |
| 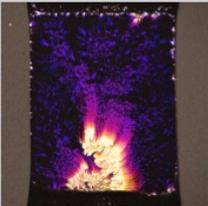 $R_F = 0.05$ $R_M = 0.01$ | 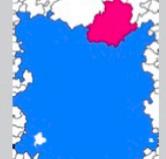 $\xi_1 = 2036$ μm $\xi_2 = 531$ μm | 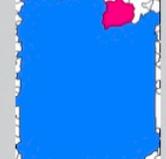 $\xi_1 = 2300$ μm $\xi_2 = 337$ μm |
| 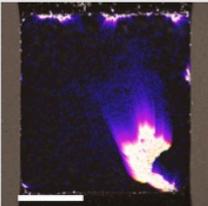 $R_F = 0.05$ $R_M = 0.01$ | 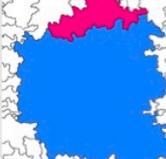 $\xi_1 = 1996$ μm $\xi_2 = 573$ μm | 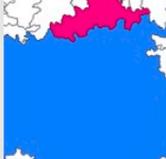 $\xi_1 = 2304$ μm $\xi_2 = 622$ μm |

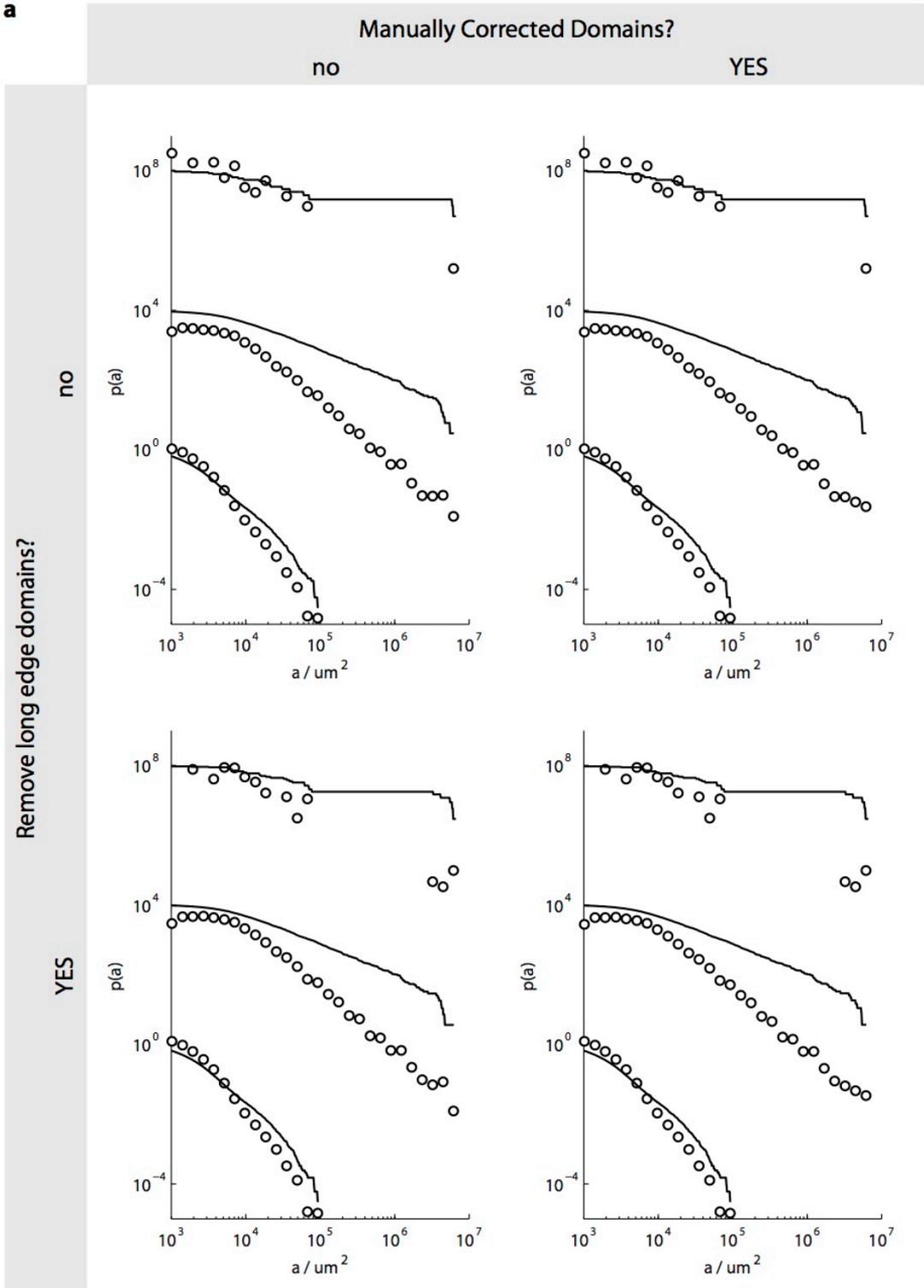